\documentclass[prb,twocolumn,groupedaddress,shownopacs,amssymb]{revtex4-2}
\usepackage{graphicx,psfrag,amsmath,amssymb}

\usepackage[usenames, dvipsnames]{color}

\usepackage{hyperref} 
\hypersetup{
    colorlinks=true,
    linkcolor=Blue,
    citecolor=Blue,
    filecolor=Blue,      
    urlcolor=Blue,
    pdftitle={Chern numbers for two-body butterfly},
    }

\begin{document}

\title{Chern numbers for the two-body Hofstadter-Hubbard butterfly}

\author{D. C. Alyuruk} 
\author{M. Iskin}
\affiliation{
Department of Physics, Ko\c{c} University, Rumelifeneri Yolu, 
34450 Sar\i yer, Istanbul, T\"urkiye
}

\date{\today}

\begin{abstract}

We analyze the two-body spectrum within the Hofstadter-Hubbard model on a square lattice 
through an exact variational ansatz and study the topological properties of its 
low-lying two-body bound-state branches. In particular we discuss how the Hofstadter-Hubbard 
butterfly of the two-body branches evolves as a function of onsite interactions and 
how to efficiently calculate their Chern numbers using the Fukui-Hatsugai-Suzuki 
approach. Our numerical results are fully consistent with the simple picture that 
appears in the strong-coupling limit, where the attraction between fermions forms 
a composite boson characterized by an effective hopping parameter and an effective
magnetic-flux ratio.

\end{abstract}
\maketitle

\section{Introduction}
\label{sec:intro}

The Hofstadter model has made profound impact on condensed-matter 
physics~\cite{hofstadter76, satija16}. 
Despite its simplicity, the intricate interplay of Aharonov-Bohm phase 
and lattice periodicity not only provides crucial insights into the behavior 
of electrons moving across a solid-state crystal in the presence of an 
external magnetic field but also shines a spotlight on one of its most 
intriguing aspects, i.e., the first Chern number. 
As long as a Bloch band remains isolated in the one-body spectrum, 
i.e., separated by finite energy gaps from the other bands, 
its associated Chern number remains constant or `protected' upon 
alterations in the magnetic field strength or the lattice potential. 
More importantly, the Chern number $C_n$ of the $n$th Bloch band 
determines the contribution of this band to the Hall 
conductivity~\cite{thouless82}. This is in such a way 
that, when the Fermi energy $\varepsilon_F$ lies within an energy gap 
labeled by $j$, the Hall conductivity is given precisely by 
$\sigma_{xy} = \sigma_j e^2/h$, where $\sigma_j = \sum_n C_n$ is a sum 
over the filled Bloch bands. Since the integer $\sigma_j$ can not change 
continuously, this result reveals that the Hall conductivity is a 
topological invariant of the system, providing insight into the 
observed robustness of the integer quantum Hall effect.
In a broader context, Chern numbers have become central to our exploration 
of topological phases of matter, illuminating phenomena as diverse as the 
quantum Hall effect, topological insulators, topological superconductors, 
and some other behavior of exotic materials under extreme 
conditions~\cite{bansil16, rachel18}. 
On the other hand, the Hubbard model is often used for probing the effects 
of strong electron-electron interactions on material properties, ranging 
from emergent phenomena such as Mott insulators, high-temperature 
superconductivity, charge-density waves, and 
magnetic ordering~\cite{arovas22}. It allows us to study how complex 
and unexpected properties emerge from the 
collective behavior of strongly-correlated electrons. 

To explore how topology influences the behavior of strongly-correlated 
electrons and vice versa, here we merge the Hofstadter and Hubbard 
models~\cite{zhai10, cocks12, repellin17, umucalilar17, zeng19, shaffer21, bartho21, shaffer22}. 
In particular, we analyze the two-body problem and formulate a two-body 
Chern number for the low-lying bound-state branches of the Hofstadter-Hubbard 
model by drawing an analogy with the Fukui-Hatsugai-Suzuki method~\cite{fukui05}. 
It is gratifying to observe that our approach successfully reproduces not 
only the anticipated butterfly spectrum but also the Chern numbers 
associated with a strongly-bound composite boson in the strong-coupling 
limit, where the composite boson is characterized by an effective 
hopping parameter and an effective magnetic-flux ratio.
This correspondence arises from the fundamental principle that the 
topological properties of a two-body branch remain unchanged as long as 
the energy spectrum remains gapped, which holds true all the way from 
the infinitely-strong-coupling limit down to a finite critical 
interaction threshold. Below this threshold, a two-body continuum begins 
to overlap, marking a transition in the system's behavior. 
We note that there are many recent works on topological aspects of 
the two-body problem in various multiband 
lattices~\cite{guo11, gorlach17, marques18, salerno18, 
lin20, zurita20, salerno20, pelegri20, zuo21, okuma23, iskin23}. 
They mostly rely on mapping the problem to an effective Hamiltonian 
for the composite bosons in the strong-coupling limit. In addition
there are some past works on the interacting butterflies in the Harper 
model~\cite{barelli96, barelli97, doh98}. Our formulation has a minor 
overlap with the existing literature, and it offers a fresh perspective 
on this long-standing problem.

The rest of the paper is organized as follows. 
In Sec.~\ref{sec:Hmodel} we introduce the usual Hofstadter model, 
and its one-body spectrum. In Sec.~\ref{sec:HHmodel} we 
introduce the Hofstadter-Hubbard model, and discuss its two-body spectrum. There we
construct the two-body butterflies in Sec.~\ref{sec:twobody}, and calculate their 
Chern numbers in Sec.~\ref{sec:chern}. The paper ends with a brief summary of our 
conclusions and an outlook in Sec.~\ref{sec:conc}.

\section{Hofstadter model}
\label{sec:Hmodel}

\begin{figure} [htb]
\includegraphics[width = 0.99\linewidth]{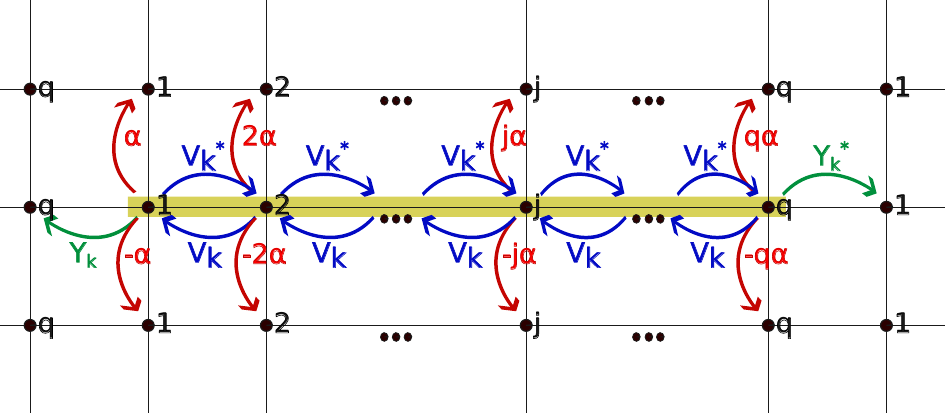}
\caption{\label{fig:ham}
Magnetic unit cell is highlighted in yellow together with its intra-unitcell 
($V_\mathbf{k}$) and inter-unitcell ($Y_\mathbf{k}$ and $Z_\mathbf{k}^j$) 
hoppings. Sublattice sites are labeled as $S \in \{1, 2, \cdots,j,\cdots q\}$.
Here the particle picks up $\pm j \alpha$ phases from upward and downward 
hoppings, respectively, leading to $Z_\mathbf{k}^j$.
}
\end{figure}

Within the tight-binding approximation, the single-particle Hamiltonian for 
a generic lattice can be written as
$
\mathcal{H}_\sigma = -\sum_{Si; S'i'} t_{Si; S'i'}^\sigma c_{S i \sigma}^\dagger c_{S' i' \sigma},
$
where the hopping parameter $t_{Si; S'i'}^\sigma$ describes tunneling of a spin 
$\sigma \in \{\uparrow, \downarrow \}$ fermion from the sublattice site $S'$ 
in the unit cell $i'$ to the sublattice site $S$ in the unit cell $i$. 
In this paper we consider a square lattice lying in the $(x, y)$ plane, and set
$
t_{Si; S'i'}^\uparrow = t_{Si; S'i'}^\downarrow = -t
$ 
for the nearest-neighbor hoppings and 0 otherwise. The presence of an external 
magnetic field 
$
\mathbf{B}(\mathbf{r}) = \boldsymbol{\nabla} \times \mathbf{A}(\mathbf{r})
$
is taken into account through the Peierls substitution
$
t \to t e^{\mathrm{i} 2\pi \phi_{Si; S'i'}}
$
with $t > 0$, where the phase factor
$
\phi_{Si; S'i'} = \frac{1}{\varphi_0} \int_{{\bf r}_{S'i'}}^{{\bf r}_{Si}} 
{\bf A}(\mathbf{r})\cdot d{\bf r}
$
takes into account the corresponding vector gauge field
~\footnote{
Since the Zeeman coupling to the spin does not have any effect on the 
two-body problem (see [38]), it is not considered in this paper.}. 
Here $\varphi_0$ is the magnetic-flux quantum and $\mathbf{r}_{Si}$ 
is the position of the site $S \in i$. 
We are interested in the original Hofstadter model~\cite{hofstadter76}, 
where a uniform magnetic field $\mathbf{B}(\mathbf{r}) = B \mathbf{\hat{z}}$
is perpendicular to a square lattice, and use the Landau gauge 
${\bf A}(\mathbf{r}) = (0, B x)$. This is such that the particle accumulates
$
\sum_\square \phi_{Si; S'i'} = Ba^2/\varphi_0 =  \alpha
$ 
uniformly after traversing around any one of the cells in the counter-clockwise 
direction, where $a$ is the lattice spacing and $\alpha$ corresponds to the number of 
magnetic-flux quanta per cell. We assume $\alpha \equiv p/q$ corresponds precisely 
to a ratio of two relatively prime numbers $p$ and $q$. In this case the 
presence of such a $B$ field leads to a (magnetic) unit cell that has $q$ sites 
in the $x$ direction, and we label its sublattice sites as $S \in \{1, 2, \cdots, q\}$.
The unit cell is illustrated in Fig.~\ref{fig:ham}.

\begin{figure} [htb]
\includegraphics[width = 0.99\linewidth]{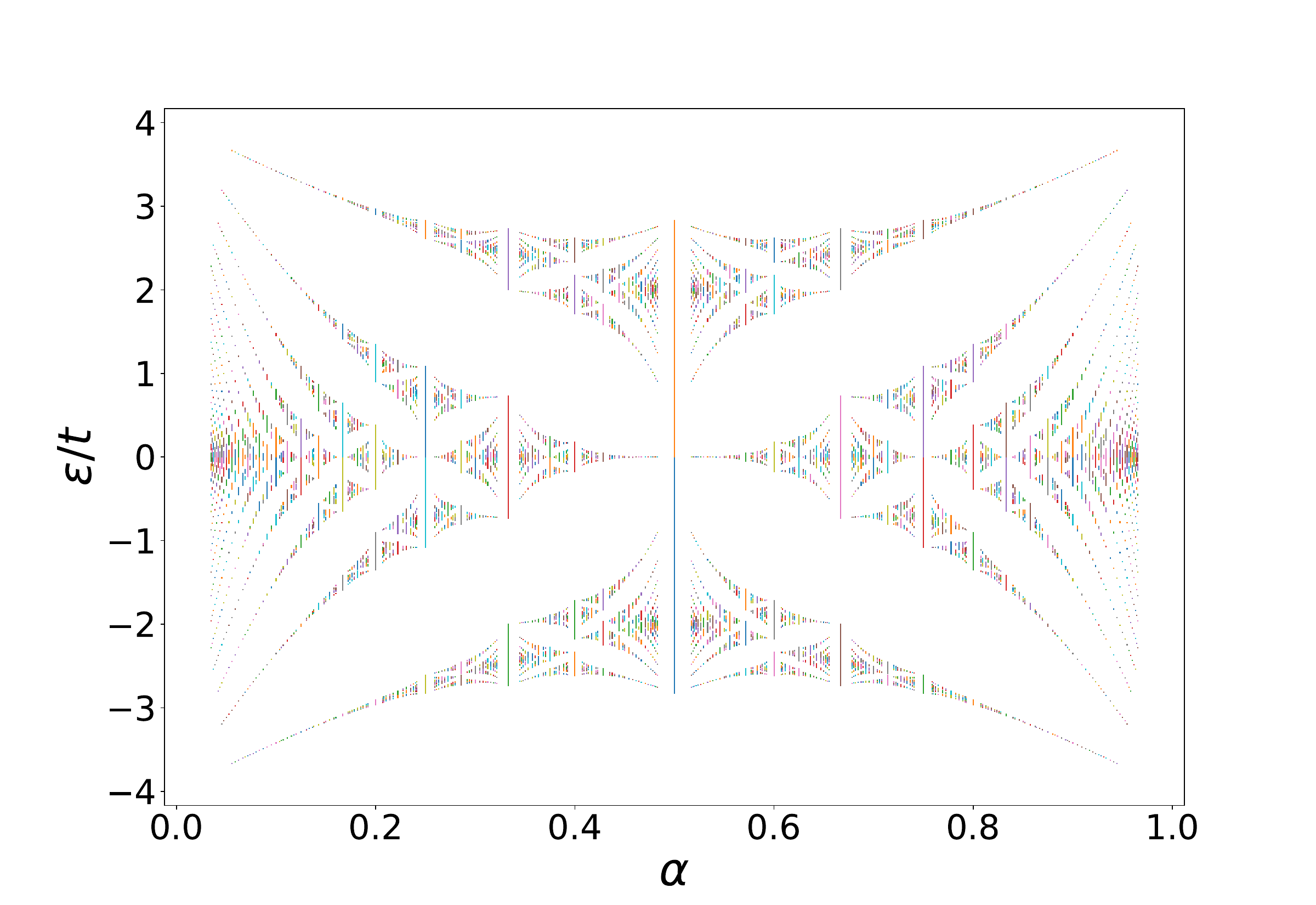}
\caption{\label{fig:Hbutterfly} 
One-body Hofstadter butterfly for the Bloch bands, where $\alpha = p/q$ is 
the number of magnetic-flux quantum per cell. Here $p$ and $q$ are relatively 
prime numbers, and $q_{max} = 30$ with all possible $p/q$ ratios.
Different Bloch bands are shown in different colors for better 
visibility, where the total bandwidths of the Bloch bands are $8t$ 
in the $\alpha \to \{0,1\}$ limits.
}
\end{figure}

Next we use the canonical transformation
$
c_{S i \sigma}^\dagger = \frac{1}{\sqrt{N_c}} \sum_\mathbf{k} 
e^{-\mathrm{i} \mathbf{k} \cdot \mathbf{r_i}} c_{S \mathbf{k} \sigma}^\dagger,
$
where $N_c$ is the number of unit cells in the system and $\mathbf{r_i}$ 
is the position of unit cell $i$, and express $\mathcal{H}_\sigma$ in the reciprocal 
space. Here $\mathbf{k} = (k_x, k_y)$ is the crystal momentum (in units of $\hbar \to 1$ 
the Plack constant) in the first magnetic Brillouin zone (MBZ), where
$
0 \le k_x < \frac{2\pi}{qa}
$
and
$
0 \le k_y < \frac{2\pi}{a}
$
are such that $\sum_\mathbf{k} 1 = N_c$. Note that the total number of cells (or 
equivalently total number of lattice sites) in the system is $N = q N_c$. 
This leads to the Bloch Hamiltonian for the Hofstadter model
$
\mathcal{H}_\sigma = \sum_{SS' {\bf k}}
h_{\bf k}^{SS'} c^{\dagger}_{S {\bf k} \sigma} c_{S' {\bf k} \sigma}
$ 
written in the sublattice basis, where the Hamiltonian matrix 
\begin{align}
\label{eq:hk}
\mathbf{h}_{\bf k} = \left(
\begin{array}{cccccc}
  Z_\mathbf{k}^1 & V_\mathbf{k} & 0 & . & 0 & Y_\mathbf{k}^* \\
  V_\mathbf{k}^* & Z_\mathbf{k}^2 & V_\mathbf{k} & 0 & . & 0 \\
  0 & \ddots & \ddots & \ddots & 0 & . \\
  . & 0 & V_\mathbf{k}^* & Z_\mathbf{k}^j & V_\mathbf{k} & 0  \\
  0 & . & 0 & \ddots & \ddots & \ddots \\
  Y_\mathbf{k} & 0 & . & 0 & V_\mathbf{k}^* & Z_\mathbf{k}^q \\
\end{array} \right)
\end{align}
is $q \times q$. Here $Z_\mathbf{k}^j  = 2t\cos(2\pi j \alpha - k_y a)$ describes 
inter-unitcell hoppings in the $y$ direction, and $V_\mathbf{k} = t$ and 
$Y_\mathbf{k} = te^{\mathrm{i} k_x q a}$ describe, respectively, the intra-unitcell 
and inter-unitcell hoppings in the $x$ direction with the periodic boundary
conditions. These processes are illustrated in Fig.~\ref{fig:ham}.
The resultant eigenvalue problem,
\begin{align}
\label{eqn:hSS}
\sum_{S'} h_\mathbf{k}^{SS'} n_{S' \mathbf{k}} = \varepsilon_{n\mathbf{k}} n_{S \mathbf{k}},
\end{align}
leads to $q$ Bloch bands in the one-body spectrum, which can be labeled as
$n \in \{1, 2, \cdots, q\}$ starting with the lowest band. 
Here $n_{S \mathbf{k}}$ is the projection of the Bloch state onto sublattice $S$.
The spectrum preserves inversion symmetry 
$
\varepsilon_{n \mathbf{k}} = \varepsilon_{n, -\mathbf{k}}
$
as a direct manifestation of the gauge invariance in a uniform flux, it has
$
\varepsilon_{n \mathbf{k}} = - \varepsilon_{q-n, -\mathbf{k}}
$
symmetry due to the particle-hole symmetry on a bipartite lattice, and it is 
mirror-symmetric 
$
\varepsilon_{n\mathbf{k}} (\alpha) = \varepsilon_{n\mathbf{k}} (1-\alpha)
$ 
around $\alpha = 1/2$ for $ 0 \le \alpha \le 1$~\cite{hofstadter76}
~\footnote{
The mirror symmetry around $\alpha = 1/2$ can be deduced from the following 
observations:
($i$) changing the direction of the magnetic field, i.e., $\alpha \to -\alpha$,
can not have any effect on the spectrum, and
($ii$) Eq.~(\ref{eq:hk}) is invariant under the addition of $2\pi j$ to 
the argument of cosine in $Z_\mathbf{k}^{-j}$.
}.
When $q$ is an even denominator, these symmetries imply
$
\varepsilon_{q/2,\mathbf{k}} = -\varepsilon_{q/2+1,\mathbf{k}},
$
so that the centrally-symmetric $n = q/2$ and $n = q/2+1$ bands touch $q$ times 
with each other at zero energy leading to $q$ Dirac cones in the MBZ.
Some of these features are visible in Fig.~\ref{fig:Hbutterfly}.

Furthermore, one of the elegant aspects of the Hofstadter model is that the 
competition between the magnetic length scale (i.e., the magnetic cyclotron 
radius) and the periodicity of the square lattice is known to produce a 
fractal pattern when bandwidths of the Bloch bands are plotted against 
$\alpha$~\cite{hofstadter76, HofstadterTools}. 
As shown in Fig.~\ref{fig:Hbutterfly}, since the shape of this pattern looks 
like the wings of a butterfly, it is usually referred to as the Hofstadter 
butterfly in the literature. Having introduced the underlying one-body problem, 
next we analyze the two-body problem.

\section{Hofstadter-Hubbard model}
\label{sec:HHmodel}

Assuming a spin-$1/2$ system, the Hamiltonian for the Hofstadter-Hubbard 
model can be written as
$
\mathcal{H} = \mathcal{H}_0 + \mathcal{H}_{\uparrow\downarrow},
$
where
$
\mathcal{H}_0 = \sum_\sigma \mathcal{H}_\sigma
$
is the hopping part, and
$
\mathcal{H}_{\uparrow\downarrow} = - U \sum_{S i} c_{S i \uparrow}^\dagger 
c_{S i \downarrow}^\dagger c_{S i \downarrow} c_{S i \uparrow}
$
takes the onsite interactions between $\uparrow$ and $\downarrow$ fermions
into account with $U \ge 0$ the strength of the attraction
~\footnote{
Our formalism is valid for both attractive ($U > 0$) and repulsive 
($U < 0$) interactions. For instance, in the latter case,
Eq.~(\ref{eqn:GK}) can be used to determine the high-lying 
two-body branches which appear at the top of the two-body spectrum. 
They are also known as the repulsively-bound doublon states in 
the literature~\cite{winkler06}. Note that such states do not appear 
in free-space models because, unlike the lattice models that 
feature a finite bandwidth, the parabolic one-body spectrum is 
not bounded from above.}. 
Using the canonical transformation given above, and upon transformation
$
c_{n \mathbf{k} \sigma}^\dagger = \sum_S n_{S \mathbf{k}} 
c_{S \mathbf{k} \sigma}^\dagger
$
to the band basis, $\mathcal{H}$ can be written as~\cite{iskin22f}
\begin{align}
\mathcal{H} &= \sum_{n \mathbf{k}} \varepsilon_{n\mathbf{k}}
c_{n \mathbf{k} \sigma}^\dagger c_{n \mathbf{k} \sigma}
- \frac{U}{N_c} \sum_{\substack{nmn'm' \\ S \mathbf{k}\mathbf{k'}\mathbf{Q}}}
n_{S \mathbf{k}}^*
m_{S, \mathbf{Q-k}}^*
\nonumber \\
&\times m_{S, \mathbf{Q-k'}}'
n_{S \mathbf{k'}}'
c_{n \mathbf{k} \uparrow}^\dagger
c_{m,\mathbf{Q-k}, \downarrow}^\dagger
c_{m',\mathbf{Q-k'}, \downarrow}
c_{n' \mathbf{k'} \uparrow}.
\label{eqn:H}
\end{align}
We emphasize that this is the exact analog of the Hofstadter-Hubbard model in 
reciprocal lattice, and it is a convenient starting point for the analysis of the 
two-body spectrum as it explicitly conserves the center-of-mass momentum 
$\mathbf{Q}$ of the incoming and outgoing particles.

\subsection{Two-body Hofstadter-Hubbard butterfly}
\label{sec:twobody}

\begin{figure} [htb]
\includegraphics[width = 0.99\linewidth]{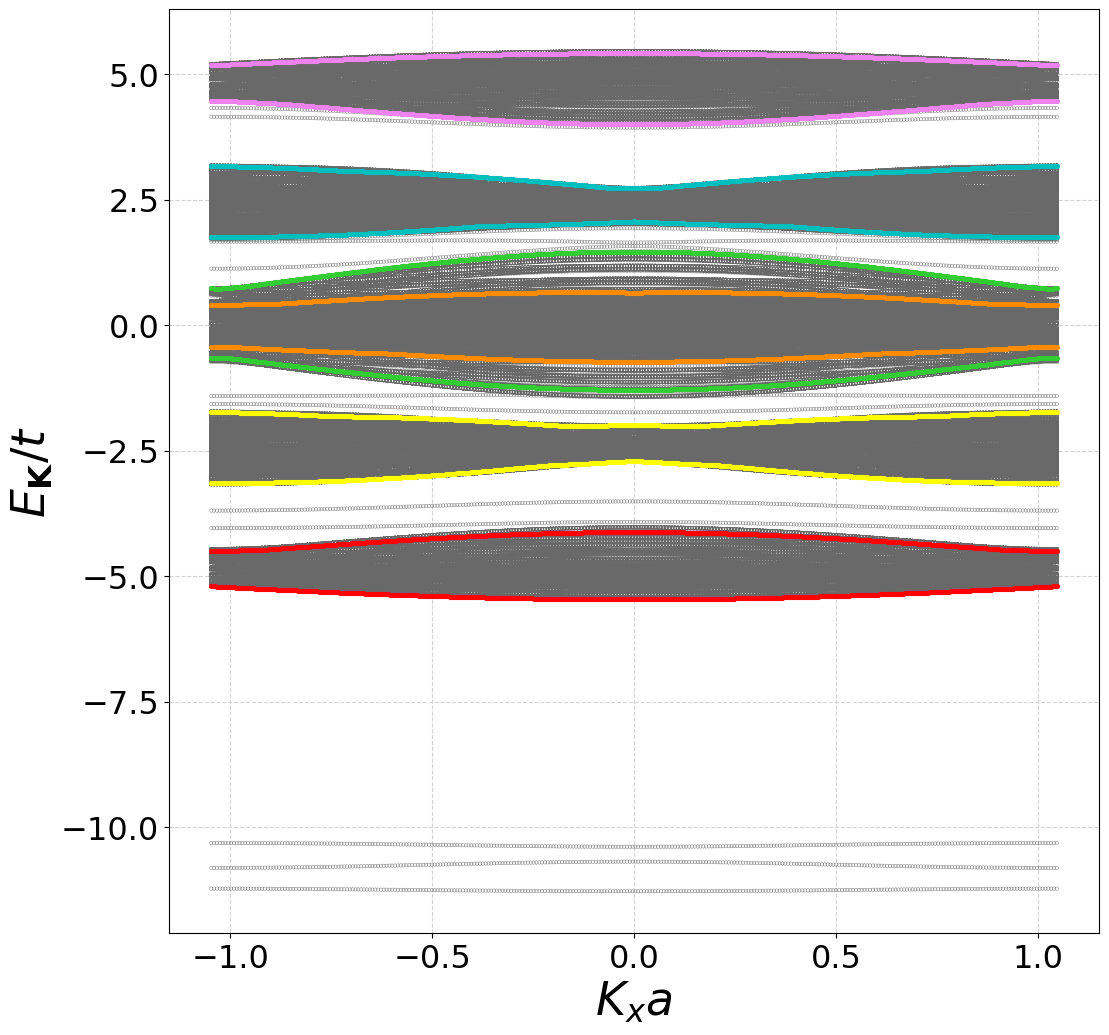}
\caption{\label{fig:13all}
Full two-body spectrum $E_\mathbf{K}$ as a function of $K_x \in \textrm{MBZ}$ 
when $K_y = 0$ and $N = 8100$. 
Gray data corresponds to solutions of Eq.~(\ref{eqn:alphanmk}) for 
$\alpha = 1/3$ when $U = 10t$.
Six pairs of colored lines are guides to the eye for the six different possible 
sets of two-body continua discussed in the text. In this paper we are interested
in the Chern numbers of the low-lying two-body branches that appear at the 
bottom of the spectrum.
}
\end{figure}

Noting that the onsite interactions allow solely a spin-singlet state, and
explicitly conserving the center-of-mass momentum $\mathbf{K}$ of the particles, 
the two-body spectrum $E_\mathbf{K}$ can be obtained exactly through the 
following ansatz~\cite{iskin21}
\begin{align}
|\psi_\mathbf{K} \rangle = \sum_{n m \mathbf{k}} \alpha_{nm}^\mathbf{k}(\mathbf{K}) 
c_{n \mathbf{k} \uparrow}^\dagger c_{m,\mathbf{K- k},\downarrow}^\dagger 
| 0 \rangle,
\end{align}
where the variational parameters satisfy
$
\alpha_{n m}^\mathbf{k} (\mathbf{K}) = \alpha_{m n}^\mathbf{K-k} (\mathbf{K}),
$
and $| 0 \rangle$ refers to the vacuum of particles. Through the functional 
minimization of
$
\langle \psi_\mathbf{K} | \mathcal{H} - E_\mathbf{K} | \psi_\mathbf{K} \rangle
$
with respect to $\alpha_{n m}^\mathbf{k} (\mathbf{K})$, we obtain a set of linear 
equations given by~\cite{iskin21}
\begin{align}
(\varepsilon_{n \mathbf{k}} &+ \varepsilon_{m, \mathbf{K - k}} 
- E_\mathbf{K} )
\alpha_{n m}^\mathbf{k} (\mathbf{K}) =
\nonumber \\
&
\frac{U}{N_c} \sum_{n' m' \mathbf{k'} S}
n_{S \mathbf{k}}^* m_{S, \mathbf{K - k}}^* 
m_{S, \mathbf{K-k'}}' n_{S \mathbf{k'}}' 
\alpha_{n' m'}^{\mathbf{k'}} (\mathbf{K}).
\label{eqn:alphanmk}
\end{align}
Thus, for any given $\mathbf{K}$, $E_\mathbf{K}$ can be determined by recasting 
Eq.~(\ref{eqn:alphanmk}) as an eigenvalue problem in the form of a 
$q^2 N_c \times q^2 N_c$ matrix. It turns out a typical two-body spectrum has 
three different sets of solutions~\cite{iskin22f}. For a given $\mathbf{K}$, there are 
($i$) $q(q+1)/2$ two-body scattering continua, ($ii$) a number of weakly-bound 
two-body bound states that always lie in between the scattering continua even 
in the $U/t \to \infty$ limit, and ($iii$) $q$ two-body bound states at the 
bottom of the spectrum which are allowed to become strongly-bound in the 
$U/t \to \infty$ limit. In this paper we are interested in formulating the 
Chern numbers of the low-lying two-body branches that appear at the bottom 
of the spectrum with a finite energy gap. 
As an illustration, we set $K_y = 0$, $\alpha = 1/3$ and $U = 10t$ in 
Fig.~\ref{fig:13all}, and present the resultant $E_\mathbf{K}$ as a function of $K_x$. 
The colored pair of lines are determined by
$
\max ( \varepsilon_{n \mathbf{k}} + \varepsilon_{m, \mathbf{K-k}} )
$
and
$
\min ( \varepsilon_{n \mathbf{k}} + \varepsilon_{m, \mathbf{K-k}} )
$
for a given $(n, m)$ combination, and six different pairs correspond to 
upper and lower edges of six possible two-body continua when $q = 3$. 
In addition there are three low-lying two-body bound-state branches 
with energies $E_\mathbf{K} \sim -U$ when $q = 3$.

\begin{figure*} [htb]
\includegraphics[width = 0.99\textwidth]{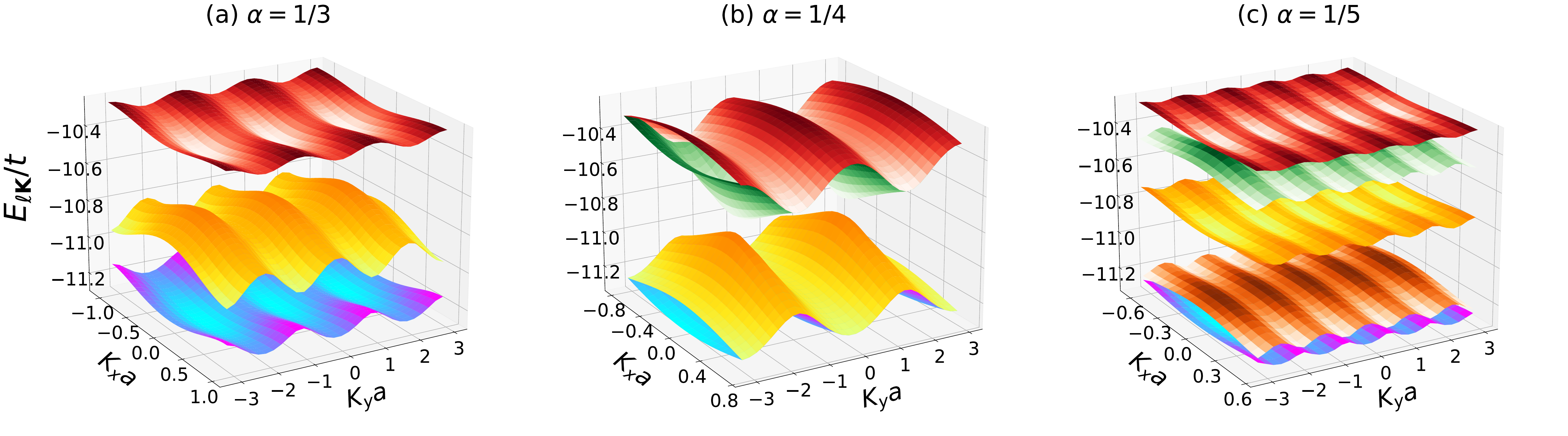}
\caption{\label{fig:bands} 
Low-lying two-body branches $E_{\ell \mathbf{K}}$ as a function of 
$\mathbf{K} \in \textrm{MBZ}$, where the index $\ell = \{1,2,\cdots,q\}$ 
starts with the lowest-lying branch.
These are self-consistent solutions of Eq.~(\ref{eqn:GK}) for (a) 
$\alpha = 1/3$, (b) $\alpha = 1/4$ and (c) $\alpha = 1/5$ when $U = 10t$.
Note that (a) is identical to the bottom of the spectrum shown in 
Fig.~\ref{fig:13all}.
}
\end{figure*}

A more powerful yet efficient way of finding the low-lying two-body branches 
of interest is as follows. By defining a new set of variational parameters,
$
\beta_{S \mathbf{K}} = \sum_{nm \mathbf{k}} \alpha_{n m}^\mathbf{k}(\mathbf{K}) 
n_{S \mathbf{k}} m_{S, \mathbf{K-k}},
$
we recast Eq.~(\ref{eqn:alphanmk}) as a nonlinear-eigenvalue problem~\cite{iskin23}
\begin{align}
\label{eqn:GK}
\mathbf{G_K} \boldsymbol{\beta}_\mathbf{K} = 0,
\end{align}
where $\mathbf{G_K}$ is a $q \times q$ Hermitian matrix in the sublattice basis 
with elements
\begin{align}
\label{eqn:GSS}
G_\mathbf{K}^{SS'} = \delta_{SS'} - \frac{U}{N_c} \sum_{n m \mathbf{k}}
\frac{n_{S' \mathbf{k}}^* m_{S', \mathbf{K-k}}^* m_{S, \mathbf{K-k}} n_{S \mathbf{k}}}
{\varepsilon_{n \mathbf{k}} + \varepsilon_{m, \mathbf{K-k}} - E_\mathbf{K}}.
\end{align}
Then we classify and distinguish solutions of Eq.~(\ref{eqn:GK}) by setting the 
eigenvalues of $\mathbf{G_K}$ to $0$ one at a time. For a given $\mathbf{K}$, 
this is equivalent to $q$ uncoupled nonlinear self-consistency equations for 
$E_\mathbf{K}$, and we keep only the lowest converging $E_\mathbf{K}$ solution 
from each equation. This leads to $q$ bound states for a given $\mathbf{K}$, 
and we label them as $E_{\ell \mathbf{K}}$ where the index 
$\ell = \{1, 2, \cdots, q\}$ starts with the lowest two-body branch. 
It turns out a particular two-body branch is 
associated with a particular eigenvalue of $\mathbf{G_K}$ for every 
$\mathbf{K} \in \textrm{MBZ}$, e.g., setting its third eigenvalue to zero 
may produce fifth branch. This approach works very well as long as the two-body
branch of interest does not overlap with a two-body continuum.
As an illustration, we set $\alpha = \{1/3, 1/4, 1/5\}$ and $U = 10t$ in 
Fig.~\ref{fig:bands}, and present the resultant $E_{\ell \mathbf{K}}$ as a 
function of $\mathbf{K}$. These results are best understood in the $U/t \to \infty$, 
where $t_b = 2t^2/U$ and $\alpha_b = 2\alpha \equiv p_b/q_b$ are, respectively, 
the effective nearest-neighbor hopping parameter and effective number of 
magnetic-flux quantum per cell for a strongly-bound pair of $\uparrow$ 
and $\downarrow$ particles. Note that, when a bound state breaks up, 
incurring a cost of binding energy $U$ in the denominator, and its $\uparrow$ 
constituent hops to a neighboring site, the $\downarrow$ partner follows suit 
and also hops to the same site. This results in a contribution of $t^2$ in 
the numerator, where the prefactor 2 takes into consideration of the 
possibility of change in the order of spins.
Here $p_b$ and $q_b$ are again relatively prime numbers, i.e., $p_b = p$ and $q_b = q/2$ 
when $q$ is even. For this reason Figs.~\ref{fig:bands}(a),~\ref{fig:bands}(b)
and~\ref{fig:bands}(c) are reminiscent of the effective Bloch bands for a 
composite boson with $\alpha_b = \{2/3,1/2,2/5\}$, below some energy offset 
(of order $-U$) determined by the binding energy. 
However, note that, since the MBZ of the composite boson is twice the 
MBZ of its constituent fermions when $q$ is an even denominator, 
$E_{\ell \mathbf{K}}$ would appear folded when plotted in the MBZ of the fermions. 
This explains the strange-looking structure of Fig.~\ref{fig:bands}(b). 

\begin{figure} [htb]
\includegraphics[width = 0.95\linewidth]{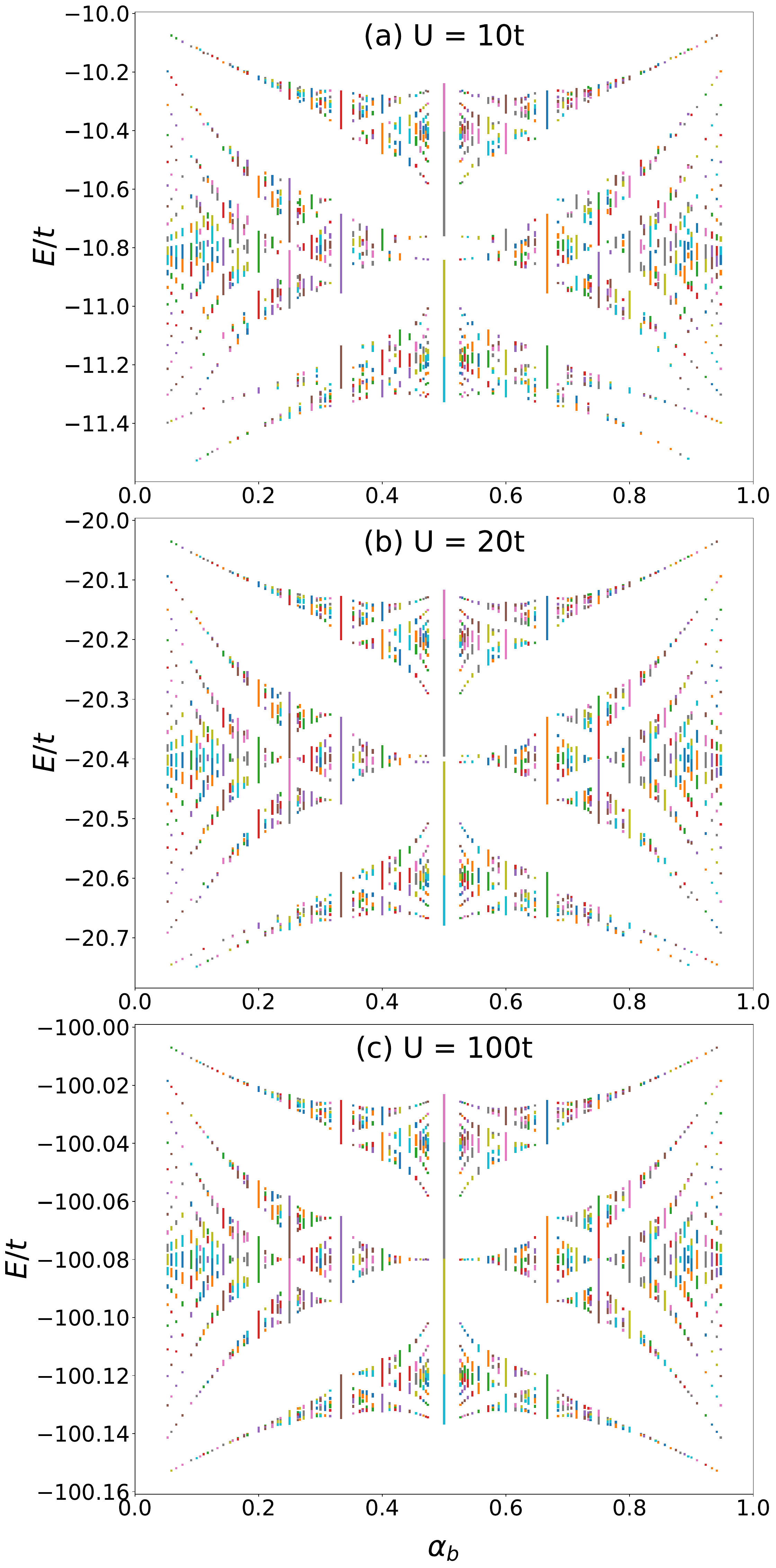}
\caption{\label{fig:HHbutterfly} 
Two-body Hofstadter-Hubbard butterfly for the low-lying two-body branches,
where $\alpha_b \equiv p_b/q_b$ is the effective number of 
magnetic-flux quantum per cell for the bound states. 
Here $p_b$ and $q_b$ are relatively prime numbers where $q_{max} = 20$ 
with all possible $p/q$ ratios.
Interaction strength $U/t$ is set to $10$ in (a), $20$ in (b) and $100$ in (c).
Different two-body branches are shown in different colors for better visibility, 
where their total bandwidths scale as $8t_b = 16t^2/U$
in the $\alpha_b \to \{0,1\}$ limits, and they are centered around $-U-8t^2/U$.
}
\end{figure}

Enchanted by the intricacies of the one-body Hofstadter butterfly, we construct and 
present the analogous Hofstadter-Hubbard butterfly for the low-lying two-body 
branches in Fig.~\ref{fig:HHbutterfly}, where different two-body branches 
are shown in different colors for better visibility. For instance, when $U = 10t$, 
one can extract the bandwidths for the $\alpha_b = \{2/3, 1/2, 2/5 \}$ ratios from 
either Fig.~\ref{fig:bands} or Fig.~\ref{fig:HHbutterfly}(a). While all of the $p/q$ 
ratios up to $q_{max} = 20$ is considered in these butterflies, in cases when different 
$p/q$ ratios are equivalent to the same $p_b/q_b$, we show the bandwidths of the 
Bloch bands for the ratio with the lowest $q$ value. This is why $\alpha_b = 1/3$ 
has three Bloch bands in Fig.~\ref{fig:HHbutterfly} even though it can come from 
either $\alpha = 2/3$ or $1/6$. In addition Fig.~\ref{fig:bands}(b) shows that 
the bandwidths of the non-isolated bands overlap in energy in the even $q$ 
case, and our coloring scheme does not distinguish these overlapping regions
as they can be included in the upper or the lower band. 
It is pleasing to see that the two-body butterfly shown in 
Fig.~\ref{fig:HHbutterfly}(c) bears resemblance to the usual Hofstadter 
butterfly in the $U/t \gg 1$ limit, where the total bandwidths of the Bloch 
bands are approximately $8t_b = 16t^2/U$ when $\alpha_b \to \{0^+, 1^-\}$. 
The two-body butterflies are centered around $-U-8t^2/U$, because,
when a bound state breaks up at a cost of $U$, one of its constituents
can hop to a neighboring site and then come back to the original site to 
recombine. This leads to an effective onsite energy $2 z_{nn} t^2/U$ for 
the pair, where the prefactor $2$ accounts for the possible hopping of
the other constituent, and $z_{nn} = 4$ is the number of 
nearest-neighbors on a square lattice~\cite{iskin23}. Having discussed 
the low-lying two-body branches, next we analyze their Chern numbers.

\subsection{Two-body Chern number}
\label{sec:chern}

As discussed in Sec.~\ref{sec:twobody}, the low-lying two-body spectrum $E_{\ell \mathbf{K}}$ 
can be determined by setting the eigenvalues of $\mathbf{G_K}$ to zero one at 
a time. Here we show that the associated eigenvectors of Eq.~(\ref{eqn:GK}), 
i.e.,
$
\boldsymbol{\beta}_{\ell \mathbf{K}} = (\beta_{1 \ell \mathbf{K}}, 
\beta_{2 \ell \mathbf{K}}, \cdots, \beta_{q \ell \mathbf{K}})^\mathrm{T}
$
in the sublattice basis with $\mathrm{T}$ the transpose, can be used to 
characterize the topology of the low-lying two-body branches~\cite{iskin23}
~\footnote{See Ref.~\cite{iskin23} for an alternative formulation of 
the two-body Chern number that is in association with a two-body 
Berry curvature. However that formulation is not exact by construction, 
it produces accurate results only in the $U/t \gg 1$ limit, and it converges 
very slowly to the expected integers in the $N_c \to \infty$ limit. 
Furthermore our current formulation is very efficient, and it yields 
correct integers above a critical $N_c$ threshold, which is typically 
very small depending on $q$ and $U/t$.}. 
For this purpose, we follow closely 
the Fukui-Hatsugai-Suzuki approach that is developed for the usual Hofstadter 
model~\cite{fukui05}, and define the so-called link variable as
\begin{equation}
\mathcal{U}_{\ell \mathbf{K}_j}^{\boldsymbol{\mu}} = \frac{\sum_S \beta_{S \ell \mathbf{K}_j}^* 
\beta_{S \ell, \mathbf{K}_j + \boldsymbol{\hat{\mu}}}}
{|\sum_S \beta_{S \ell \mathbf{K}_j}^* 
\beta_{S \ell, \mathbf{K}_j + \boldsymbol{\hat{\mu}}}|}
\end{equation}
for the $\ell$th branch, where 
$
\mathbf{K}_j = \big( \frac{2\pi}{q_b N_x a} j_x, \frac{2\pi}{N_y a} j_y \big)
$
with $j_x = \{0, 1, 2, \cdots, N_x-1\}$ and $j_y = \{0, 1, 2, \cdots, N_y-1\}$
denotes the position of a lattice point in the effective MBZ. Furthermore,
$
\boldsymbol{\hat{\mu}} \in \{\mathbf{\hat{x}}, \mathbf{\hat{y}}\}
$ 
is a vector pointing along the $K_x$ or $K_y$ axis in the effective MBZ, where
$
\mathbf{\hat{x}} = \big( \frac{2\pi}{q_b N_x a}, 0 \big)
$
and
$
\mathbf{\hat{y}} = \big( 0, \frac{2\pi}{N_y a} \big).
$
Note that $N_c = N_x N_y$ is the number of primitive unit cells, and we choose
$q_b N_x = N_y$.
In addition the eigenvectors $\boldsymbol{\beta}_{\ell \mathbf{K}}$ are periodic 
in the reciprocal space by construction, where
$
\beta_{S \ell \mathbf{K}_j} 
= \beta_{S \ell, \mathbf{K}_j + N_x \mathbf{\hat{x}}}
= \beta_{S \ell, \mathbf{K}_j + N_y \mathbf{\hat{y}}}.
$ 
Then we define the so-called field strength by
\begin{equation}
F_{\ell \mathbf{K}_j} = \ln\big[ \mathcal{U}_{\ell \mathbf{K}_j}^\mathbf{x} 
\mathcal{U}_{\ell, \mathbf{K}_j + \mathbf{\hat{x}}}^\mathbf{y} 
(\mathcal{U}_{\ell, \mathbf{K}_j + \mathbf{\hat{y}}}^\mathbf{x})^{-1}
(\mathcal{U}_{\ell \mathbf{K}_j}^\mathbf{y})^{-1} \big]
\end{equation}
for the $\ell$th branch, within the principal branch of the logarithm where 
$
-\pi < \frac{1}{\mathrm{i}} F_{\ell \mathbf{K}_j} \le \pi
$ 
counts the accumulated phase after traversing the cell around the point 
$\mathbf{K}_j$ (which corresponds to the left-lower corner of the cell) 
in the counter-clockwise direction.
This leads to the Chern number of the $\ell$th low-lying two-body branch 
as~\cite{fukui05}
\begin{equation}
C_\ell = \frac{1}{2 \pi \mathrm{i}} \sum_{j_x j_y} F_{\ell \mathbf{K}_j},
\end{equation}
where the summation covers the effective MBZ. Note that $C_\ell$ can only be defined 
when the $\ell$th branch is well separated from other states, i.e., when it 
satisfies the gap-opening condition
$
|E_{\ell \mathbf{K}} - E_{\ell \pm 1, \mathbf{K}}| \ne 0
$
for all $\mathbf{K} \in \textrm{MBZ}$ states in the sufficiently large $U/t$ 
regime. 

Our approach works very well and reproduces the anticipated $C_\ell$ above 
a critical $U$ threshold (which is approximately of the order of the total 
bandwidth $W_\alpha$ of the Bloch bands for a given $\alpha$) as long as the 
two-body branch of interest does not overlap with another two-body branch or 
a two-body continuum. For instance, when $\alpha = p/q$ 
in the usual Hofstadter model, the Chern number of the $j$th energy gap 
in the Bloch spectrum is known to satisfy the Diophantine equation 
$\sigma_j \equiv s j$ in$\mod q$, where $s$ is the modular inverse of $p$, 
i.e., $s p \equiv 1$ in$\mod q$~\cite{avron14, thouless82}.
Thus, $\sigma_j = \sum_{n \leq j} C_n$, where $C_n$ is the Chern number of the 
$n$th Bloch band with the index $n = \{1,2,\cdots, q\}$ starting from the 
lowest-lying one. The Diophantine equation leaves a$\mod q$ ambiguity in $\sigma_j$, 
except for the $0$th and $q$th gap in which case $\sigma_0 = \sigma_q = 0$
corresponds, respectively, to a trivial particle vacuum and a trivial band 
insulator. This ambiguity was resolved for the Hofstadter model on the 
rectangular lattice~\cite{avron14, thouless82}, leading to the constraint
$
\sigma_j \in \left[ 1 - \frac{q}{2}, \frac{q}{2} -1 \right]
$
when $q$ is an even denominator, and to the constraint
$
\sigma_j \in \left[ - \frac{q - 1}{2}, \frac{q-1 }{2} \right] 
$
when $q$ is an odd denominator. For example, since $s = 1$ when $\alpha = 1/3$, 
we find $\sigma_j = \{0,1,2,0\}$ from the Diophantine equation, leading to 
$\sigma_j \to \{0,1,-1,0\}$ for the $j$th gap in the constraining interval and 
to $C_n \to \{+1,-2,+1\}$ for the Bloch bands. Similarly, since $s = 2$ when 
$\alpha = 2/3$, we find $\sigma_j = \{0,2,1,0\}$ from the Diophantine equation, 
leading to $\sigma_j \to \{0,-1,+1,0\}$ in the constraining interval and to 
$C_n \to \{-1,+2,-1\}$ for the Bloch bands. Similarly, since $s = 3$ when 
$\alpha = 2/5$, we find $\sigma_j = \{0,3,1,4,2,0\}$ from the Diophantine equation, 
leading to $\sigma_j \to \{0,-2,+1,-1,+2,0\}$ in the constraining interval 
and to $C_n \to \{-2,+3,-2,+3,-2\}$ for the Bloch bands. These are in perfect
agreement with our numerical $C_\ell$ values, where the effective flux ratios 
are $\alpha_b = \{1/3, 2/3, 2/5\}$ for the bound states when 
$\alpha = \{2/3, 1/3, 1/5\}$ . 
We also checked many other flux ratios, e.g.,  $C_\ell = \{-3,+4,-3,+4,-3,+4,-3 \}$ 
when $\alpha = 1/7$, and $C_\ell = \{-1,-1,2,2,-1,-1 \}$ when 
$\alpha_\uparrow = 1/2$ is different from $\alpha_\downarrow = 1/3$
~\footnote{
We note that, in the general case when the $\uparrow$ and $\downarrow$ 
fermions have distinct one-body spectra, i.e., when 
$
\mathbf{h}_{\bf k}^\uparrow \ne \mathbf{h}_{\bf k}^\downarrow,
$
one needs to replace the Bloch bands 
$
\varepsilon_{n \mathbf{k}} \to \varepsilon_{n \mathbf{k} \uparrow}
$
and 
$
\varepsilon_{m, \mathbf{K-k}} \to \varepsilon_{m, \mathbf{K-k}, \downarrow}
$
and the Bloch factors
$
n_{S \mathbf{k}} \to n_{S \mathbf{k} \uparrow}
$ 
and
$
m_{S, \mathbf{K-k}} \to m_{S, \mathbf{K-k}, \downarrow},
$
accordingly in Eqs.~(\ref{eqn:alphanmk}) and~(\ref{eqn:GSS})~\cite{iskin21}.}.
Note that the middle branches $\ell = \{3,4\}$ are not energetically isolated 
and touch each other in the latter case since 
$
\alpha_b = \alpha_\uparrow + \alpha_\downarrow = 5/6
$ 
has an even denominator, i.e., $C_3$ and $C_4$ are not well defined.

\section{Conclusion}
\label{sec:conc}

In summary, here we analyzed the two-body problem within the Hofstadter-Hubbard 
model, with a particular focus on its low-lying two-body bound-state branches. 
In particular we studied evolution of their two-body Hofstadter-Hubbard butterfly 
as a function of the interaction strength $U$, and formulated their Chern numbers 
$C_\ell$ in an efficient way by making an analogy with the Fukui-Hatsugai-Suzuki 
method~\cite{fukui05}. 
Our numerical results at finite $U$ are in perfect agreement with the expected 
Chern numbers associated with a composite boson in the $U/t \to \infty$ limit, 
where $t_b = 2t^2/U$ and $\alpha_b = 2\alpha$ are, 
respectively, the effective hopping parameter and effective magnetic-flux ratio. 
This is because the topological nature of a two-body branch cannot change 
as long as its spectrum remains gapped, which turns out to be 
the case down to a critical threshold $U \sim W_\alpha$ determined by the 
total bandwidth $W_\alpha$ of the Bloch bands for a given $\alpha$, below which 
the low-lying two-body branches start overlapping with a two-body continuum. 

Recent studies have highlighted the significance of the Chern number of a Bloch 
band within topological band theory~\cite{bansil16}, where it serves as a powerful 
tool for characterizing and comprehending the topological characteristics of 
electronic band structures in various materials. For instance it plays a pivotal 
role in classifying topological phases, elucidating the quantization of the Hall 
conductance, and predicting the emergence of novel electronic states~\cite{rachel18}. 
Similarly, the Chern number of a two-body bound-state branch may find some 
potential applications and utility in certain physical phenomena, and its fate 
will be determined in time. It is worth emphasizing that our formalism is 
quite generic and valid not only for attractive ($U > 0$) and repulsive 
($U < 0$) onsite interactions, but also it is readily applicable to all 
sorts of lattice geometries. As an outlook we expect the single-particle 
bulk-boundary correspondence to apply to the two-body topological phase as well. 
For instance, similar to the recent results on the interacting Haldane 
model~\cite{salerno18, iskin23}, one can verify that the two-body Chern 
numbers of the interacting Hofstadter model are also in agreement with 
the chirality of the edge states through an exact diagonalization with 
open boundary conditions~\cite{hatsugai93}. In addition we expect the 
two-body analogue of the conventional Hall conductance to be 
$
\bar{\sigma}_{xy} = \bar{\sigma}_j \frac{\bar{e}^2} {h},
$
where $\bar{\sigma}_j = \sum_\ell C_\ell$ and $\bar{e} = 2e$ 
is the effective charge of the pairs. Finally it is possible to extend 
our approach to finite-range interactions which is underway.

\begin{acknowledgments}
The authors acknowledge funding from T{\"U}B{\.I}TAK.
\end{acknowledgments}

\bibliography{refs}

\begin{thebibliography}{40}%
\makeatletter
\providecommand \@ifxundefined [1]{%
 \@ifx{#1\undefined}
}%
\providecommand \@ifnum [1]{%
 \ifnum #1\expandafter \@firstoftwo
 \else \expandafter \@secondoftwo
 \fi
}%
\providecommand \@ifx [1]{%
 \ifx #1\expandafter \@firstoftwo
 \else \expandafter \@secondoftwo
 \fi
}%
\providecommand \natexlab [1]{#1}%
\providecommand \enquote  [1]{``#1''}%
\providecommand \bibnamefont  [1]{#1}%
\providecommand \bibfnamefont [1]{#1}%
\providecommand \citenamefont [1]{#1}%
\providecommand \href@noop [0]{\@secondoftwo}%
\providecommand \href [0]{\begingroup \@sanitize@url \@href}%
\providecommand \@href[1]{\@@startlink{#1}\@@href}%
\providecommand \@@href[1]{\endgroup#1\@@endlink}%
\providecommand \@sanitize@url [0]{\catcode `\\12\catcode `\$12\catcode `\&12\catcode `\#12\catcode `\^12\catcode `\_12\catcode `\%12\relax}%
\providecommand \@@startlink[1]{}%
\providecommand \@@endlink[0]{}%
\providecommand \url  [0]{\begingroup\@sanitize@url \@url }%
\providecommand \@url [1]{\endgroup\@href {#1}{\urlprefix }}%
\providecommand \urlprefix  [0]{URL }%
\providecommand \Eprint [0]{\href }%
\providecommand \doibase [0]{https://doi.org/}%
\providecommand \selectlanguage [0]{\@gobble}%
\providecommand \bibinfo  [0]{\@secondoftwo}%
\providecommand \bibfield  [0]{\@secondoftwo}%
\providecommand \translation [1]{[#1]}%
\providecommand \BibitemOpen [0]{}%
\providecommand \bibitemStop [0]{}%
\providecommand \bibitemNoStop [0]{.\EOS\space}%
\providecommand \EOS [0]{\spacefactor3000\relax}%
\providecommand \BibitemShut  [1]{\csname bibitem#1\endcsname}%
\let\auto@bib@innerbib\@empty
\bibitem [{\citenamefont {Hofstadter}(1976)}]{hofstadter76}%
  \BibitemOpen
  \bibfield  {author} {\bibinfo {author} {\bibfnamefont {D.~R.}\ \bibnamefont {Hofstadter}},\ }\bibfield  {title} {\bibinfo {title} {Energy levels and wave functions of bloch electrons in rational and irrational magnetic fields},\ }\href {https://doi.org/10.1103/PhysRevB.14.2239} {\bibfield  {journal} {\bibinfo  {journal} {Phys. Rev. B}\ }\textbf {\bibinfo {volume} {14}},\ \bibinfo {pages} {2239} (\bibinfo {year} {1976})}\BibitemShut {NoStop}%
\bibitem [{\citenamefont {Satija}(2016)}]{satija16}%
  \BibitemOpen
  \bibfield  {author} {\bibinfo {author} {\bibfnamefont {I.~I.}\ \bibnamefont {Satija}},\ }\href@noop {} {\emph {\bibinfo {title} {The Butterfly in the Quantum World: The story of the most fascinating quantum fractal}}}\ (\bibinfo  {publisher} {Morgan \& Claypool Publishers},\ \bibinfo {year} {2016})\BibitemShut {NoStop}%
\bibitem [{\citenamefont {Thouless}\ \emph {et~al.}(1982)\citenamefont {Thouless}, \citenamefont {Kohmoto}, \citenamefont {Nightingale},\ and\ \citenamefont {den Nijs}}]{thouless82}%
  \BibitemOpen
  \bibfield  {author} {\bibinfo {author} {\bibfnamefont {D.~J.}\ \bibnamefont {Thouless}}, \bibinfo {author} {\bibfnamefont {M.}~\bibnamefont {Kohmoto}}, \bibinfo {author} {\bibfnamefont {M.~P.}\ \bibnamefont {Nightingale}},\ and\ \bibinfo {author} {\bibfnamefont {M.}~\bibnamefont {den Nijs}},\ }\bibfield  {title} {\bibinfo {title} {Quantized {Hall} conductance in a two-dimensional periodic potential},\ }\href {https://doi.org/10.1103/PhysRevLett.49.405} {\bibfield  {journal} {\bibinfo  {journal} {Phys. Rev. Lett.}\ }\textbf {\bibinfo {volume} {49}},\ \bibinfo {pages} {405} (\bibinfo {year} {1982})}\BibitemShut {NoStop}%
\bibitem [{\citenamefont {Bansil}\ \emph {et~al.}(2016)\citenamefont {Bansil}, \citenamefont {Lin},\ and\ \citenamefont {Das}}]{bansil16}%
  \BibitemOpen
  \bibfield  {author} {\bibinfo {author} {\bibfnamefont {A.}~\bibnamefont {Bansil}}, \bibinfo {author} {\bibfnamefont {H.}~\bibnamefont {Lin}},\ and\ \bibinfo {author} {\bibfnamefont {T.}~\bibnamefont {Das}},\ }\bibfield  {title} {\bibinfo {title} {Colloquium: Topological band theory},\ }\href {https://doi.org/10.1103/RevModPhys.88.021004} {\bibfield  {journal} {\bibinfo  {journal} {Rev. Mod. Phys.}\ }\textbf {\bibinfo {volume} {88}},\ \bibinfo {pages} {021004} (\bibinfo {year} {2016})}\BibitemShut {NoStop}%
\bibitem [{\citenamefont {Rachel}(2018)}]{rachel18}%
  \BibitemOpen
  \bibfield  {author} {\bibinfo {author} {\bibfnamefont {S.}~\bibnamefont {Rachel}},\ }\bibfield  {title} {\bibinfo {title} {Interacting topological insulators: a review},\ }\href {https://doi.org/10.1088/1361-6633/aad6a6} {\bibfield  {journal} {\bibinfo  {journal} {Reports on Progress in Physics}\ }\textbf {\bibinfo {volume} {81}},\ \bibinfo {pages} {116501} (\bibinfo {year} {2018})}\BibitemShut {NoStop}%
\bibitem [{\citenamefont {Arovas}\ \emph {et~al.}(2022)\citenamefont {Arovas}, \citenamefont {Berg}, \citenamefont {Kivelson},\ and\ \citenamefont {Raghu}}]{arovas22}%
  \BibitemOpen
  \bibfield  {author} {\bibinfo {author} {\bibfnamefont {D.~P.}\ \bibnamefont {Arovas}}, \bibinfo {author} {\bibfnamefont {E.}~\bibnamefont {Berg}}, \bibinfo {author} {\bibfnamefont {S.~A.}\ \bibnamefont {Kivelson}},\ and\ \bibinfo {author} {\bibfnamefont {S.}~\bibnamefont {Raghu}},\ }\bibfield  {title} {\bibinfo {title} {The {Hubbard} model},\ }\href@noop {} {\bibfield  {journal} {\bibinfo  {journal} {Annual review of condensed matter physics}\ }\textbf {\bibinfo {volume} {13}},\ \bibinfo {pages} {239} (\bibinfo {year} {2022})}\BibitemShut {NoStop}%
\bibitem [{\citenamefont {Zhai}\ \emph {et~al.}(2010)\citenamefont {Zhai}, \citenamefont {Umucal{\i}lar},\ and\ \citenamefont {Oktel}}]{zhai10}%
  \BibitemOpen
  \bibfield  {author} {\bibinfo {author} {\bibfnamefont {H.}~\bibnamefont {Zhai}}, \bibinfo {author} {\bibfnamefont {R.~O.}\ \bibnamefont {Umucal{\i}lar}},\ and\ \bibinfo {author} {\bibfnamefont {M.~O.}\ \bibnamefont {Oktel}},\ }\bibfield  {title} {\bibinfo {title} {Pairing and vortex lattices for interacting fermions in optical lattices with a large magnetic field},\ }\href {https://doi.org/10.1103/PhysRevLett.104.145301} {\bibfield  {journal} {\bibinfo  {journal} {Phys. Rev. Lett.}\ }\textbf {\bibinfo {volume} {104}},\ \bibinfo {pages} {145301} (\bibinfo {year} {2010})}\BibitemShut {NoStop}%
\bibitem [{\citenamefont {Cocks}\ \emph {et~al.}(2012)\citenamefont {Cocks}, \citenamefont {Orth}, \citenamefont {Rachel}, \citenamefont {Buchhold}, \citenamefont {Le~Hur},\ and\ \citenamefont {Hofstetter}}]{cocks12}%
  \BibitemOpen
  \bibfield  {author} {\bibinfo {author} {\bibfnamefont {D.}~\bibnamefont {Cocks}}, \bibinfo {author} {\bibfnamefont {P.~P.}\ \bibnamefont {Orth}}, \bibinfo {author} {\bibfnamefont {S.}~\bibnamefont {Rachel}}, \bibinfo {author} {\bibfnamefont {M.}~\bibnamefont {Buchhold}}, \bibinfo {author} {\bibfnamefont {K.}~\bibnamefont {Le~Hur}},\ and\ \bibinfo {author} {\bibfnamefont {W.}~\bibnamefont {Hofstetter}},\ }\bibfield  {title} {\bibinfo {title} {Time-reversal-invariant {H}ofstadter-{H}ubbard model with ultracold fermions},\ }\href {https://doi.org/10.1103/PhysRevLett.109.205303} {\bibfield  {journal} {\bibinfo  {journal} {Phys. Rev. Lett.}\ }\textbf {\bibinfo {volume} {109}},\ \bibinfo {pages} {205303} (\bibinfo {year} {2012})}\BibitemShut {NoStop}%
\bibitem [{\citenamefont {Repellin}\ \emph {et~al.}(2017)\citenamefont {Repellin}, \citenamefont {Yefsah},\ and\ \citenamefont {Sterdyniak}}]{repellin17}%
  \BibitemOpen
  \bibfield  {author} {\bibinfo {author} {\bibfnamefont {C.}~\bibnamefont {Repellin}}, \bibinfo {author} {\bibfnamefont {T.}~\bibnamefont {Yefsah}},\ and\ \bibinfo {author} {\bibfnamefont {A.}~\bibnamefont {Sterdyniak}},\ }\bibfield  {title} {\bibinfo {title} {Creating a bosonic fractional quantum {Hall} state by pairing fermions},\ }\href {https://doi.org/10.1103/PhysRevB.96.161111} {\bibfield  {journal} {\bibinfo  {journal} {Phys. Rev. B}\ }\textbf {\bibinfo {volume} {96}},\ \bibinfo {pages} {161111} (\bibinfo {year} {2017})}\BibitemShut {NoStop}%
\bibitem [{\citenamefont {Umucal{\i}lar}\ and\ \citenamefont {Iskin}(2017)}]{umucalilar17}%
  \BibitemOpen
  \bibfield  {author} {\bibinfo {author} {\bibfnamefont {R.~O.}\ \bibnamefont {Umucal{\i}lar}}\ and\ \bibinfo {author} {\bibfnamefont {M.}~\bibnamefont {Iskin}},\ }\bibfield  {title} {\bibinfo {title} {{BCS} theory of time-reversal-symmetric {Hofstadter-Hubbard} model},\ }\href {https://doi.org/10.1103/PhysRevLett.119.085301} {\bibfield  {journal} {\bibinfo  {journal} {Phys. Rev. Lett.}\ }\textbf {\bibinfo {volume} {119}},\ \bibinfo {pages} {085301} (\bibinfo {year} {2017})}\BibitemShut {NoStop}%
\bibitem [{\citenamefont {Zeng}\ \emph {et~al.}(2019)\citenamefont {Zeng}, \citenamefont {Stanescu}, \citenamefont {Zhang}, \citenamefont {Scarola},\ and\ \citenamefont {Tewari}}]{zeng19}%
  \BibitemOpen
  \bibfield  {author} {\bibinfo {author} {\bibfnamefont {C.}~\bibnamefont {Zeng}}, \bibinfo {author} {\bibfnamefont {T.~D.}\ \bibnamefont {Stanescu}}, \bibinfo {author} {\bibfnamefont {C.}~\bibnamefont {Zhang}}, \bibinfo {author} {\bibfnamefont {V.~W.}\ \bibnamefont {Scarola}},\ and\ \bibinfo {author} {\bibfnamefont {S.}~\bibnamefont {Tewari}},\ }\bibfield  {title} {\bibinfo {title} {Majorana corner modes with solitons in an attractive {Hubbard-Hofstadter} model of cold atom optical lattices},\ }\href {https://doi.org/10.1103/PhysRevLett.123.060402} {\bibfield  {journal} {\bibinfo  {journal} {Phys. Rev. Lett.}\ }\textbf {\bibinfo {volume} {123}},\ \bibinfo {pages} {060402} (\bibinfo {year} {2019})}\BibitemShut {NoStop}%
\bibitem [{\citenamefont {Shaffer}\ \emph {et~al.}(2021)\citenamefont {Shaffer}, \citenamefont {Wang},\ and\ \citenamefont {Santos}}]{shaffer21}%
  \BibitemOpen
  \bibfield  {author} {\bibinfo {author} {\bibfnamefont {D.}~\bibnamefont {Shaffer}}, \bibinfo {author} {\bibfnamefont {J.}~\bibnamefont {Wang}},\ and\ \bibinfo {author} {\bibfnamefont {L.~H.}\ \bibnamefont {Santos}},\ }\bibfield  {title} {\bibinfo {title} {Theory of {H}ofstadter superconductors},\ }\href {https://doi.org/10.1103/PhysRevB.104.184501} {\bibfield  {journal} {\bibinfo  {journal} {Phys. Rev. B}\ }\textbf {\bibinfo {volume} {104}},\ \bibinfo {pages} {184501} (\bibinfo {year} {2021})}\BibitemShut {NoStop}%
\bibitem [{\citenamefont {Andrews}\ \emph {et~al.}(2021)\citenamefont {Andrews}, \citenamefont {Neupert},\ and\ \citenamefont {M\"oller}}]{bartho21}%
  \BibitemOpen
  \bibfield  {author} {\bibinfo {author} {\bibfnamefont {B.}~\bibnamefont {Andrews}}, \bibinfo {author} {\bibfnamefont {T.}~\bibnamefont {Neupert}},\ and\ \bibinfo {author} {\bibfnamefont {G.}~\bibnamefont {M\"oller}},\ }\bibfield  {title} {\bibinfo {title} {Stability, phase transitions, and numerical breakdown of fractional chern insulators in higher {Chern} bands of the {Hofstadter} model},\ }\href {https://doi.org/10.1103/PhysRevB.104.125107} {\bibfield  {journal} {\bibinfo  {journal} {Phys. Rev. B}\ }\textbf {\bibinfo {volume} {104}},\ \bibinfo {pages} {125107} (\bibinfo {year} {2021})}\BibitemShut {NoStop}%
\bibitem [{\citenamefont {Shaffer}\ \emph {et~al.}(2022)\citenamefont {Shaffer}, \citenamefont {Wang},\ and\ \citenamefont {Santos}}]{shaffer22}%
  \BibitemOpen
  \bibfield  {author} {\bibinfo {author} {\bibfnamefont {D.}~\bibnamefont {Shaffer}}, \bibinfo {author} {\bibfnamefont {J.}~\bibnamefont {Wang}},\ and\ \bibinfo {author} {\bibfnamefont {L.~H.}\ \bibnamefont {Santos}},\ }\bibfield  {title} {\bibinfo {title} {Unconventional self-similar {Hofstadter} superconductivity from repulsive interactions},\ }\href@noop {} {\bibfield  {journal} {\bibinfo  {journal} {Nature Communications}\ }\textbf {\bibinfo {volume} {13}},\ \bibinfo {pages} {7785} (\bibinfo {year} {2022})}\BibitemShut {NoStop}%
\bibitem [{\citenamefont {Fukui}\ \emph {et~al.}(2005)\citenamefont {Fukui}, \citenamefont {Hatsugai},\ and\ \citenamefont {Suzuki}}]{fukui05}%
  \BibitemOpen
  \bibfield  {author} {\bibinfo {author} {\bibfnamefont {T.}~\bibnamefont {Fukui}}, \bibinfo {author} {\bibfnamefont {Y.}~\bibnamefont {Hatsugai}},\ and\ \bibinfo {author} {\bibfnamefont {H.}~\bibnamefont {Suzuki}},\ }\bibfield  {title} {\bibinfo {title} {Chern numbers in discretized {B}rillouin zone: efficient method of computing (spin) {H}all conductances},\ }\href@noop {} {\bibfield  {journal} {\bibinfo  {journal} {Journal of the Physical Society of Japan}\ }\textbf {\bibinfo {volume} {74}},\ \bibinfo {pages} {1674} (\bibinfo {year} {2005})}\BibitemShut {NoStop}%
\bibitem [{\citenamefont {Guo}\ and\ \citenamefont {Shen}(2011)}]{guo11}%
  \BibitemOpen
  \bibfield  {author} {\bibinfo {author} {\bibfnamefont {H.}~\bibnamefont {Guo}}\ and\ \bibinfo {author} {\bibfnamefont {S.-Q.}\ \bibnamefont {Shen}},\ }\bibfield  {title} {\bibinfo {title} {Topological phase in a one-dimensional interacting fermion system},\ }\href {https://doi.org/10.1103/PhysRevB.84.195107} {\bibfield  {journal} {\bibinfo  {journal} {Phys. Rev. B}\ }\textbf {\bibinfo {volume} {84}},\ \bibinfo {pages} {195107} (\bibinfo {year} {2011})}\BibitemShut {NoStop}%
\bibitem [{\citenamefont {Gorlach}\ and\ \citenamefont {Poddubny}(2017)}]{gorlach17}%
  \BibitemOpen
  \bibfield  {author} {\bibinfo {author} {\bibfnamefont {M.~A.}\ \bibnamefont {Gorlach}}\ and\ \bibinfo {author} {\bibfnamefont {A.~N.}\ \bibnamefont {Poddubny}},\ }\bibfield  {title} {\bibinfo {title} {Topological edge states of bound photon pairs},\ }\href {https://doi.org/10.1103/PhysRevA.95.053866} {\bibfield  {journal} {\bibinfo  {journal} {Phys. Rev. A}\ }\textbf {\bibinfo {volume} {95}},\ \bibinfo {pages} {053866} (\bibinfo {year} {2017})}\BibitemShut {NoStop}%
\bibitem [{\citenamefont {Marques}\ and\ \citenamefont {Dias}(2018)}]{marques18}%
  \BibitemOpen
  \bibfield  {author} {\bibinfo {author} {\bibfnamefont {A.}~\bibnamefont {Marques}}\ and\ \bibinfo {author} {\bibfnamefont {R.}~\bibnamefont {Dias}},\ }\bibfield  {title} {\bibinfo {title} {Topological bound states in interacting {S}u--{S}chrieffer--{H}eeger rings},\ }\href@noop {} {\bibfield  {journal} {\bibinfo  {journal} {Journal of Physics: Condensed Matter}\ }\textbf {\bibinfo {volume} {30}},\ \bibinfo {pages} {305601} (\bibinfo {year} {2018})}\BibitemShut {NoStop}%
\bibitem [{\citenamefont {Salerno}\ \emph {et~al.}(2018)\citenamefont {Salerno}, \citenamefont {Di~Liberto}, \citenamefont {Menotti},\ and\ \citenamefont {Carusotto}}]{salerno18}%
  \BibitemOpen
  \bibfield  {author} {\bibinfo {author} {\bibfnamefont {G.}~\bibnamefont {Salerno}}, \bibinfo {author} {\bibfnamefont {M.}~\bibnamefont {Di~Liberto}}, \bibinfo {author} {\bibfnamefont {C.}~\bibnamefont {Menotti}},\ and\ \bibinfo {author} {\bibfnamefont {I.}~\bibnamefont {Carusotto}},\ }\bibfield  {title} {\bibinfo {title} {Topological two-body bound states in the interacting {H}aldane model},\ }\href {https://doi.org/10.1103/PhysRevA.97.013637} {\bibfield  {journal} {\bibinfo  {journal} {Phys. Rev. A}\ }\textbf {\bibinfo {volume} {97}},\ \bibinfo {pages} {013637} (\bibinfo {year} {2018})}\BibitemShut {NoStop}%
\bibitem [{\citenamefont {Lin}\ \emph {et~al.}(2020)\citenamefont {Lin}, \citenamefont {Ke},\ and\ \citenamefont {Lee}}]{lin20}%
  \BibitemOpen
  \bibfield  {author} {\bibinfo {author} {\bibfnamefont {L.}~\bibnamefont {Lin}}, \bibinfo {author} {\bibfnamefont {Y.}~\bibnamefont {Ke}},\ and\ \bibinfo {author} {\bibfnamefont {C.}~\bibnamefont {Lee}},\ }\bibfield  {title} {\bibinfo {title} {Interaction-induced topological bound states and {T}houless pumping in a one-dimensional optical lattice},\ }\href {https://doi.org/10.1103/PhysRevA.101.023620} {\bibfield  {journal} {\bibinfo  {journal} {Phys. Rev. A}\ }\textbf {\bibinfo {volume} {101}},\ \bibinfo {pages} {023620} (\bibinfo {year} {2020})}\BibitemShut {NoStop}%
\bibitem [{\citenamefont {Zurita}\ \emph {et~al.}(2020)\citenamefont {Zurita}, \citenamefont {Creffield},\ and\ \citenamefont {Platero}}]{zurita20}%
  \BibitemOpen
  \bibfield  {author} {\bibinfo {author} {\bibfnamefont {J.}~\bibnamefont {Zurita}}, \bibinfo {author} {\bibfnamefont {C.~E.}\ \bibnamefont {Creffield}},\ and\ \bibinfo {author} {\bibfnamefont {G.}~\bibnamefont {Platero}},\ }\bibfield  {title} {\bibinfo {title} {Topology and interactions in the photonic {C}reutz and {C}reutz-{H}ubbard ladders},\ }\href@noop {} {\bibfield  {journal} {\bibinfo  {journal} {Advanced Quantum Technologies}\ }\textbf {\bibinfo {volume} {3}},\ \bibinfo {pages} {1900105} (\bibinfo {year} {2020})}\BibitemShut {NoStop}%
\bibitem [{\citenamefont {Salerno}\ \emph {et~al.}(2020)\citenamefont {Salerno}, \citenamefont {Palumbo}, \citenamefont {Goldman},\ and\ \citenamefont {Di~Liberto}}]{salerno20}%
  \BibitemOpen
  \bibfield  {author} {\bibinfo {author} {\bibfnamefont {G.}~\bibnamefont {Salerno}}, \bibinfo {author} {\bibfnamefont {G.}~\bibnamefont {Palumbo}}, \bibinfo {author} {\bibfnamefont {N.}~\bibnamefont {Goldman}},\ and\ \bibinfo {author} {\bibfnamefont {M.}~\bibnamefont {Di~Liberto}},\ }\bibfield  {title} {\bibinfo {title} {Interaction-induced lattices for bound states: Designing flat bands, quantized pumps, and higher-order topological insulators for doublons},\ }\href {https://doi.org/10.1103/PhysRevResearch.2.013348} {\bibfield  {journal} {\bibinfo  {journal} {Phys. Rev. Res.}\ }\textbf {\bibinfo {volume} {2}},\ \bibinfo {pages} {013348} (\bibinfo {year} {2020})}\BibitemShut {NoStop}%
\bibitem [{\citenamefont {Pelegr\'{\i}}\ \emph {et~al.}(2020)\citenamefont {Pelegr\'{\i}}, \citenamefont {Marques}, \citenamefont {Ahufinger}, \citenamefont {Mompart},\ and\ \citenamefont {Dias}}]{pelegri20}%
  \BibitemOpen
  \bibfield  {author} {\bibinfo {author} {\bibfnamefont {G.}~\bibnamefont {Pelegr\'{\i}}}, \bibinfo {author} {\bibfnamefont {A.~M.}\ \bibnamefont {Marques}}, \bibinfo {author} {\bibfnamefont {V.}~\bibnamefont {Ahufinger}}, \bibinfo {author} {\bibfnamefont {J.}~\bibnamefont {Mompart}},\ and\ \bibinfo {author} {\bibfnamefont {R.~G.}\ \bibnamefont {Dias}},\ }\bibfield  {title} {\bibinfo {title} {Interaction-induced topological properties of two bosons in flat-band systems},\ }\href {https://doi.org/10.1103/PhysRevResearch.2.033267} {\bibfield  {journal} {\bibinfo  {journal} {Phys. Rev. Res.}\ }\textbf {\bibinfo {volume} {2}},\ \bibinfo {pages} {033267} (\bibinfo {year} {2020})}\BibitemShut {NoStop}%
\bibitem [{\citenamefont {Zuo}\ \emph {et~al.}(2021)\citenamefont {Zuo}, \citenamefont {Benalcazar}, \citenamefont {Liu},\ and\ \citenamefont {Liu}}]{zuo21}%
  \BibitemOpen
  \bibfield  {author} {\bibinfo {author} {\bibfnamefont {Z.-W.}\ \bibnamefont {Zuo}}, \bibinfo {author} {\bibfnamefont {W.~A.}\ \bibnamefont {Benalcazar}}, \bibinfo {author} {\bibfnamefont {Y.}~\bibnamefont {Liu}},\ and\ \bibinfo {author} {\bibfnamefont {C.-X.}\ \bibnamefont {Liu}},\ }\bibfield  {title} {\bibinfo {title} {Topological phases of the dimerized {Hofstadter} butterfly},\ }\href@noop {} {\bibfield  {journal} {\bibinfo  {journal} {Journal of Physics D: Applied Physics}\ }\textbf {\bibinfo {volume} {54}},\ \bibinfo {pages} {414004} (\bibinfo {year} {2021})}\BibitemShut {NoStop}%
\bibitem [{\citenamefont {Okuma}\ and\ \citenamefont {Mizoguchi}(2023)}]{okuma23}%
  \BibitemOpen
  \bibfield  {author} {\bibinfo {author} {\bibfnamefont {N.}~\bibnamefont {Okuma}}\ and\ \bibinfo {author} {\bibfnamefont {T.}~\bibnamefont {Mizoguchi}},\ }\bibfield  {title} {\bibinfo {title} {Relationship between two-particle topology and fractional {C}hern insulator},\ }\href {https://doi.org/10.1103/PhysRevResearch.5.013112} {\bibfield  {journal} {\bibinfo  {journal} {Phys. Rev. Res.}\ }\textbf {\bibinfo {volume} {5}},\ \bibinfo {pages} {013112} (\bibinfo {year} {2023})}\BibitemShut {NoStop}%
\bibitem [{\citenamefont {Iskin}(2023)}]{iskin23}%
  \BibitemOpen
  \bibfield  {author} {\bibinfo {author} {\bibfnamefont {M.}~\bibnamefont {Iskin}},\ }\bibfield  {title} {\bibinfo {title} {Topological two-body bands in a multiband {Hubbard} model},\ }\href {https://doi.org/10.1103/PhysRevA.107.053323} {\bibfield  {journal} {\bibinfo  {journal} {Phys. Rev. A}\ }\textbf {\bibinfo {volume} {107}},\ \bibinfo {pages} {053323} (\bibinfo {year} {2023})}\BibitemShut {NoStop}%
\bibitem [{\citenamefont {Barelli}\ \emph {et~al.}(1996)\citenamefont {Barelli}, \citenamefont {Bellissard}, \citenamefont {Jacquod},\ and\ \citenamefont {Shepelyansky}}]{barelli96}%
  \BibitemOpen
  \bibfield  {author} {\bibinfo {author} {\bibfnamefont {A.}~\bibnamefont {Barelli}}, \bibinfo {author} {\bibfnamefont {J.}~\bibnamefont {Bellissard}}, \bibinfo {author} {\bibfnamefont {P.}~\bibnamefont {Jacquod}},\ and\ \bibinfo {author} {\bibfnamefont {D.~L.}\ \bibnamefont {Shepelyansky}},\ }\bibfield  {title} {\bibinfo {title} {Double butterfly spectrum for two interacting particles in the {Harper} model},\ }\href {https://doi.org/10.1103/PhysRevLett.77.4752} {\bibfield  {journal} {\bibinfo  {journal} {Phys. Rev. Lett.}\ }\textbf {\bibinfo {volume} {77}},\ \bibinfo {pages} {4752} (\bibinfo {year} {1996})}\BibitemShut {NoStop}%
\bibitem [{\citenamefont {Barelli}\ \emph {et~al.}(1997)\citenamefont {Barelli}, \citenamefont {Bellissard}, \citenamefont {Jacquod},\ and\ \citenamefont {Shepelyansky}}]{barelli97}%
  \BibitemOpen
  \bibfield  {author} {\bibinfo {author} {\bibfnamefont {A.}~\bibnamefont {Barelli}}, \bibinfo {author} {\bibfnamefont {J.}~\bibnamefont {Bellissard}}, \bibinfo {author} {\bibfnamefont {P.}~\bibnamefont {Jacquod}},\ and\ \bibinfo {author} {\bibfnamefont {D.~L.}\ \bibnamefont {Shepelyansky}},\ }\bibfield  {title} {\bibinfo {title} {Two interacting {Hofstadter} butterflies},\ }\href {https://doi.org/10.1103/PhysRevB.55.9524} {\bibfield  {journal} {\bibinfo  {journal} {Phys. Rev. B}\ }\textbf {\bibinfo {volume} {55}},\ \bibinfo {pages} {9524} (\bibinfo {year} {1997})}\BibitemShut {NoStop}%
\bibitem [{\citenamefont {Doh}\ and\ \citenamefont {Salk}(1998)}]{doh98}%
  \BibitemOpen
  \bibfield  {author} {\bibinfo {author} {\bibfnamefont {H.}~\bibnamefont {Doh}}\ and\ \bibinfo {author} {\bibfnamefont {S.-H.~S.}\ \bibnamefont {Salk}},\ }\bibfield  {title} {\bibinfo {title} {Effects of electron correlations on the {Hofstadter} spectrum},\ }\href {https://doi.org/10.1103/PhysRevB.57.1312} {\bibfield  {journal} {\bibinfo  {journal} {Phys. Rev. B}\ }\textbf {\bibinfo {volume} {57}},\ \bibinfo {pages} {1312} (\bibinfo {year} {1998})}\BibitemShut {NoStop}%
\bibitem [{Note1()}]{Note1}%
  \BibitemOpen
  \bibinfo {note} {Since the Zeeman coupling to the spin does not have any effect on the two-body problem (see [38]), it is not considered in this paper.}\BibitemShut {Stop}%
\bibitem [{Note2()}]{Note2}%
  \BibitemOpen
  \bibinfo {note} {The mirror symmetry around $\alpha = 1/2$ can be deduced from the following observations: ($i$) changing the direction of the magnetic field, i.e., $\alpha \to -\alpha $, can not have any effect on the spectrum, and ($ii$) Eq.~(\ref {eq:hk}) is invariant under the addition of $2\pi j$ to the argument of cosine in $Z_\protect \mathbf {k}^{-j}$.}\BibitemShut {Stop}%
\bibitem [{\citenamefont {Andrews}(2023)}]{HofstadterTools}%
  \BibitemOpen
  \bibfield  {author} {\bibinfo {author} {\bibfnamefont {B.}~\bibnamefont {Andrews}},\ }\href@noop {} {\bibinfo {title} {Hofstadter{T}ools: A python package for analyzing the {Hofstadter} model}} (\bibinfo {year} {2023}),\ \Eprint {https://arxiv.org/abs/2311.18726} {arXiv:2311.18726 [cond-mat.mes-hall]} \BibitemShut {NoStop}%
\bibitem [{Note3()}]{Note3}%
  \BibitemOpen
  \bibinfo {note} {Our formalism is valid for both attractive ($U > 0$) and repulsive ($U < 0$) interactions. For instance, in the latter case, Eq.~(\ref {eqn:GK}) can be used to determine the high-lying two-body branches which appear at the top of the two-body spectrum. They are also known as the repulsively-bound doublon states in the literature~\cite {winkler06}. Note that such states do not appear in free-space models because, unlike the lattice models that feature a finite bandwidth, the parabolic one-body spectrum is not bounded from above.}\BibitemShut {Stop}%
\bibitem [{\citenamefont {Iskin}\ and\ \citenamefont {Kele\ifmmode~\mbox{\c{s}}\else \c{s}\fi{}}(2022)}]{iskin22f}%
  \BibitemOpen
  \bibfield  {author} {\bibinfo {author} {\bibfnamefont {M.}~\bibnamefont {Iskin}}\ and\ \bibinfo {author} {\bibfnamefont {A.}~\bibnamefont {Kele\ifmmode~\mbox{\c{s}}\else \c{s}\fi{}}},\ }\bibfield  {title} {\bibinfo {title} {Stability of $({N}+1)$-body fermion clusters in a multiband {H}ubbard model},\ }\href {https://doi.org/10.1103/PhysRevA.106.033304} {\bibfield  {journal} {\bibinfo  {journal} {Phys. Rev. A}\ }\textbf {\bibinfo {volume} {106}},\ \bibinfo {pages} {033304} (\bibinfo {year} {2022})}\BibitemShut {NoStop}%
\bibitem [{\citenamefont {Iskin}(2021)}]{iskin21}%
  \BibitemOpen
  \bibfield  {author} {\bibinfo {author} {\bibfnamefont {M.}~\bibnamefont {Iskin}},\ }\bibfield  {title} {\bibinfo {title} {Two-body problem in a multiband lattice and the role of quantum geometry},\ }\href {https://doi.org/10.1103/physreva.103.053311} {\bibfield  {journal} {\bibinfo  {journal} {Phys. Rev. A}\ }\textbf {\bibinfo {volume} {103}},\ \bibinfo {pages} {053311} (\bibinfo {year} {2021})}\BibitemShut {NoStop}%
\bibitem [{Note4()}]{Note4}%
  \BibitemOpen
  \bibinfo {note} {See Ref.~\cite {iskin23} for an alternative formulation of the two-body Chern number that is in association with a two-body Berry curvature. However that formulation is not exact by construction, it produces accurate results only in the $U/t \gg 1$ limit, and it converges very slowly to the expected integers in the $N_c \to \infty $ limit. Furthermore our current formulation is very efficient, and it yields correct integers above a critical $N_c$ threshold, which is typically very small depending on $q$ and $U/t$.}\BibitemShut {Stop}%
\bibitem [{\citenamefont {Avron}\ \emph {et~al.}(2014)\citenamefont {Avron}, \citenamefont {Kenneth},\ and\ \citenamefont {Yehoshua}}]{avron14}%
  \BibitemOpen
  \bibfield  {author} {\bibinfo {author} {\bibfnamefont {J.}~\bibnamefont {Avron}}, \bibinfo {author} {\bibfnamefont {O.}~\bibnamefont {Kenneth}},\ and\ \bibinfo {author} {\bibfnamefont {G.}~\bibnamefont {Yehoshua}},\ }\bibfield  {title} {\bibinfo {title} {A study of the ambiguity in the solutions to the {Diophantine} equation for {Chern} numbers},\ }\href@noop {} {\bibfield  {journal} {\bibinfo  {journal} {Journal of Physics A: Mathematical and Theoretical}\ }\textbf {\bibinfo {volume} {47}},\ \bibinfo {pages} {185202} (\bibinfo {year} {2014})}\BibitemShut {NoStop}%
\bibitem [{Note5()}]{Note5}%
  \BibitemOpen
  \bibinfo {note} {We note that, in the general case when the $\uparrow $ and $\downarrow $ fermions have distinct one-body spectra, i.e., when $ \protect \mathbf {h}_{\protect \bf k}^\uparrow \protect \ne \protect \mathbf {h}_{\protect \bf k}^\downarrow , $ one needs to replace the Bloch bands $ \varepsilon _{n \protect \mathbf {k}} \to \varepsilon _{n \protect \mathbf {k} \uparrow } $ and $ \varepsilon _{m, \protect \mathbf {K-k}} \to \varepsilon _{m, \protect \mathbf {K-k}, \downarrow } $ and the Bloch factors $ n_{S \protect \mathbf {k}} \to n_{S \protect \mathbf {k} \uparrow } $ and $ m_{S, \protect \mathbf {K-k}} \to m_{S, \protect \mathbf {K-k}, \downarrow }, $ accordingly in Eqs.~(\ref {eqn:alphanmk}) and~(\ref {eqn:GSS})~\cite {iskin21}.}\BibitemShut {Stop}%
\bibitem [{\citenamefont {Hatsugai}(1993)}]{hatsugai93}%
  \BibitemOpen
  \bibfield  {author} {\bibinfo {author} {\bibfnamefont {Y.}~\bibnamefont {Hatsugai}},\ }\bibfield  {title} {\bibinfo {title} {Chern number and edge states in the integer quantum {Hall} effect},\ }\href {https://doi.org/10.1103/PhysRevLett.71.3697} {\bibfield  {journal} {\bibinfo  {journal} {Phys. Rev. Lett.}\ }\textbf {\bibinfo {volume} {71}},\ \bibinfo {pages} {3697} (\bibinfo {year} {1993})}\BibitemShut {NoStop}%
\bibitem [{\citenamefont {Winkler}\ \emph {et~al.}(2006)\citenamefont {Winkler}, \citenamefont {Thalhammer}, \citenamefont {Lang}, \citenamefont {Grimm}, \citenamefont {Hecker~Denschlag}, \citenamefont {Daley}, \citenamefont {Kantian}, \citenamefont {B{\"u}chler},\ and\ \citenamefont {Zoller}}]{winkler06}%
  \BibitemOpen
  \bibfield  {author} {\bibinfo {author} {\bibfnamefont {K.}~\bibnamefont {Winkler}}, \bibinfo {author} {\bibfnamefont {G.}~\bibnamefont {Thalhammer}}, \bibinfo {author} {\bibfnamefont {F.}~\bibnamefont {Lang}}, \bibinfo {author} {\bibfnamefont {R.}~\bibnamefont {Grimm}}, \bibinfo {author} {\bibfnamefont {J.}~\bibnamefont {Hecker~Denschlag}}, \bibinfo {author} {\bibfnamefont {A.}~\bibnamefont {Daley}}, \bibinfo {author} {\bibfnamefont {A.}~\bibnamefont {Kantian}}, \bibinfo {author} {\bibfnamefont {H.}~\bibnamefont {B{\"u}chler}},\ and\ \bibinfo {author} {\bibfnamefont {P.}~\bibnamefont {Zoller}},\ }\bibfield  {title} {\bibinfo {title} {Repulsively bound atom pairs in an optical lattice},\ }\href@noop {} {\bibfield  {journal} {\bibinfo  {journal} {Nature}\ }\textbf {\bibinfo {volume} {441}},\ \bibinfo {pages} {853} (\bibinfo {year} {2006})}\BibitemShut {NoStop}%
\end{thebibliography}%

\end{document}